\documentclass[english,aps,twocolumn,pra]{revtex4}

\usepackage[T1]{fontenc}

\usepackage{graphicx}
\usepackage{epstopdf}
\usepackage{amssymb}
\usepackage{amsmath}
\usepackage{subfigure}
\usepackage[urlcolor=blue,hyperindex,colorlinks,bookmarks=true,linkcolor=black,citecolor=black]{hyperref}
\usepackage[normalem]{ulem}
\usepackage[abs]{overpic}
\usepackage{color}
\usepackage{rotating}

\usepackage[english]{babel}
\usepackage{blindtext,tikz}
\usetikzlibrary{calc}

\newcommand{\be}{\begin{equation}}
\newcommand{\ee}{\end{equation}}
\newcommand{\bea}{\begin{align}}
\newcommand{\eea}{\end{align}}

\newcommand{\ket}[1]{\left|#1\right\rangle}
\newcommand{\bra}[1]{\left\langle#1\right|}
\newcommand{\braket}[2]{\left\langle#1\right|\left.#2\right\rangle}

\newcommand{\abs}[1]{\lvert#1\rvert}

\newcommand{\al}{\alpha}
\newcommand{\ww}{\omega}
\newcommand{\eps}{\epsilon}

\begin{document}

\title{Entanglement Generated by the Dispersive Interaction: The Dressed Coherent State}

\author{Luke C.G. Govia}
\email[Electronic address: ]{lcggovia@lusi.uni-sb.de}
\affiliation{Theoretical Physics, Universit\"{a}t des Saarlandes, Campus, 66123 Saarbr\"{u}cken, Germany}
\author{Frank K. Wilhelm}
\affiliation{Theoretical Physics, Universit\"{a}t des Saarlandes, Campus, 66123 Saarbr\"{u}cken, Germany}

\begin{abstract}
In the dispersive regime of qubit-cavity coupling, classical cavity drive populates the cavity, but leaves the qubit state unaffected. However, the dispersive Hamiltonian is derived after both a frame transformation and an approximation. Therefore, to connect to external experimental devices, the inverse frame transformation from the dispersive frame back to the lab frame is necessary. In this work, we show that in the lab frame the system is best described by an entangled state known as the dressed coherent state, and thus even in the dispersive regime, entanglement is generated between the qubit and the cavity. Also, we show that further qubit evolution depends on both the amplitude and phase of the dressed coherent state, and use the dressed coherent state to calculate the measurement contrast of a recently developed dispersive readout protocol.
\end{abstract}

\maketitle

The interaction between a two level system (TLS) and quantized electromagnetic radiation has been studied extensively since the beginnings of quantum mechanics, with much effort devoted to the study of physical systems described by the Jaynes-Cummings Hamiltonian \cite{Jaynes63}. Over the last few decades the fields of cavity quantum electrodynamics (CQED) and more recently circuit quantum electrodynamics (cQED) have significantly developed, allowing for the exploration of the Jaynes-Cummings interaction in a wide range of parameter regimes and physical systems. In particular in cQED, both the strong coupling regime ($g\gg \kappa,\gamma$, first achieved in Rydberg atoms \cite{Meschede:1985kq}) and the strong dispersive regime ($\chi\gg\kappa,\gamma$) have been reached within the last decade \cite{Wallraff04}. In cQED, a superconducting qubit serves as the TLS, while the quantized electromagnetic fields are microwaves in either a strip-line resonator or 3D microwave cavity.

Contemporary experiments in cQED often work in the strong dispersive regime, where the qubit and microwave cavity are off resonance, and the Jaynes-Cummings interaction reduces to an effective second order shift in system eigen-energies. In this regime, a wide range of quantum information protocols has been demonstrated \cite{Devoret:2013xy}, including quantum teleportation \cite{Steffen2013}, entanglement generation by measurement and feedback \cite{Riste2013,Chow:2014fk}, non-classical microwave state generation \cite{Kirchmair:2013uq}, and error correction by stabilization measurements \cite{Kelly:2015ef}.

When an empty electromagnetic cavity is driven by classical radiation, the state of the cavity is described quantum mechanically by the coherent state $\ket{\al}$, where the complex amplitude $\al$ depends on the strength and length of the classical drive. In the dispersive regime of qubit-cavity coupling, when a classical cavity drive is applied the state of the joint system is typically described by the product state $a\ket{g}\ket{\al_g} +b\ket{e}\ket{\al_e}$, with no qubit-cavity entanglement generated if the qubit is not initially in a superposition state.

What is often overlooked is that the state $a\ket{g}\ket{\al_g} +b\ket{e}\ket{\al_e}$ is an accurate description of the joint system state under the dispersive approximation, which involves a frame transformation to the {\it dispersive frame}, and thus this state is not an accurate description of the qubit-cavity system in the lab frame of the experiment. In this manuscript, we will show that in the lab frame of the experiment a more accurate description of the joint state is the dressed coherent state $\overline{\ket{g/e,\al}}$ \cite{Sete:2013rt}. Unlike the description in the dispersive frame, the dressed coherent state of the lab frame is entangled, even if the qubit is not initially in a superposition state. This has profound implications on the future evolution of the system, and we will show that future qubit evolution is dependent on both the amplitude and phase of the dressed coherent state. Similar effects have previously been studied for a driven qubit-cavity system where the qubit and cavity are resonant \cite{Peano:2010yq,Peano:2010ys}.

This manuscript is organized as follows: in section \ref{sec:Sys} we describe the physical system of interest and the unitary frames we will be working in; in section \ref{sec:Dressed} we define the dressed coherent state and give analytical and numerical evidence that this is the state created after the coupled system is driven classically through the cavity degrees of freedom; in section \ref{sec:discuss} we discuss applications of the dressed coherent state to quantum information protocols; finally, in section \ref{sec:conc} we make concluding remarks.

\section{The Physical System}
\label{sec:Sys}
We consider a qubit and cavity coupled via the Jaynes-Cummings interaction, described by the lab frame Hamiltonian
\begin{align}
\hat{H} = \ww_{\rm c}\hat{a}^{\dagger}\hat{a} - \frac{\ww_{\rm q}}{2}\hat{\sigma}_z + g\left(\hat{\sigma}^{-}\hat{a}^{\dagger}+\hat{\sigma}^{+}\hat{a}\right),
\label{eqn:JC}
\end{align}
where $\hat{a}$ and $\hat{a}^{\dagger}$ are the usual bosonic annihilation and creation operators, $\hat{\sigma}_z$ is the Pauli matrix whose eigenstates are the qubit logical states, $\hat{\sigma}^{\pm}$ are the qubit raising and lowering operators, $\ww_{\rm c/q}$ are the cavity and qubit frequencies, $g$ is the Jaynes-Cummings coupling strength, and we set $\hbar = 1$ from here on. The eigenbasis for this Hamiltonian is given by the dressed states \cite{Haroche06}
\begin{align}
&\overline{\ket{g,n}} = \cos{\theta_{n}}\ket{g,n} - \sin{\theta_{n}}\ket{e,n-1},\label{eqn:Geig} \\
&\overline{\ket{e,n-1}} = \cos{\theta_{n}}\ket{e,n-1} + \sin{\theta_{n}}\ket{g,n}, \label{eqn:Eeig} 
\end{align}
where the unbarred kets are the eigenstates of the uncoupled system, and the mixing angle $\theta_n$ is given by the relation
\be
\theta_n = \frac{1}{2}\arctan{\left(2\lambda\sqrt{n}\right)},
\ee
where $\lambda = g/\Delta$ with $\Delta = \ww_{\rm q} - \ww_{\rm c }$ the cavity-qubit detuning. We work in the dispersive regime, defined by $\abs{\lambda} \ll 1$. In this regime the mixing angle is well approximated as $\theta_n \approx \lambda\sqrt{n}$, provided $n \ll n_{\rm crit}$, where $n_{\rm crit}$ is the critical photon number at which the approximation breaks down due to the product $\sqrt{n}\lambda$ approaching unity \cite{Blais2004}. The dressed states then reduce to
\begin{align}
&\overline{\ket{g,n}} = \cos\left({\lambda\sqrt{n}}\right)\ket{g,n} - \sin\left({\lambda\sqrt{n}}\right)\ket{e,n-1},\label{eqn:aGeig} \\
&\overline{\ket{e,n-1}} = \cos\left({\lambda\sqrt{n}}\right)\ket{e,n-1} + \sin\left({\lambda\sqrt{n}}\right)\ket{g,n}, \label{eqn:aEeig} 
\end{align}
In this regime, it is then possible to transform to the {\it dispersive frame} by applying the unitary rotation $\hat{U}_{\rm D} = \exp\left\{\lambda\left(\hat{\sigma}^{+}\hat{a}-\hat{\sigma}^{-}\hat{a}^{\dagger}\right)\right\}$, and keeping terms up to first order in the dispersive shift $\chi = g^2/\Delta$. The result of this procedure is the dispersive frame Hamiltonian
\be
\hat{H}_{\rm D} = \ww_{\rm c}\hat{a}^{\dagger}\hat{a} - \frac{\ww_{\rm q}+\chi}{2}\hat{\sigma}_z -\chi\hat{\sigma}_z \hat{a}^{\dagger}\hat{a},
\label{eqn:Disp}
\ee
which we emphasize is not in the lab frame, but in the dispersive frame defined by $\hat{U}_{\rm D}$. To highlight the in-equivalence of the two frames, we note the identities (which will be useful later)
\begin{align}
&\hat{U}^{\dagger}_{\rm D}\ket{g,n} = \overline{\ket{g,n}}, \label{eqn:gUDdag} \\
&\hat{U}^{\dagger}_{\rm D}\ket{e,n} = \overline{\ket{e,n}},
\end{align}
where the dressed states are given by equations (\ref{eqn:aGeig}) and (\ref{eqn:aEeig}).

\section{The Dressed Coherent State}
\label{sec:Dressed}

\subsection{Analytic Derivation}
\label{sec:Cdrive}

We are interested in the effect on the full system of a classical drive applied to the cavity, as described in the lab frame and to lowest order in the dispersive frame by the Hamiltonian
\be
\hat{H}_{\rm d}(t) = 2\cos(\ww_{\rm d}t)\left(\eps\hat{a}+\eps^*\hat{a}^{\dagger}\right),
\label{eqn:CavDri}
\ee
where $\ww_{\rm d}$ and $\eps$ are the frequency and complex amplitude of the drive respectively. In the interaction frame of the system Hamiltonian $\hat{H}_{\rm D}$ of equation (\ref{eqn:Disp}) (which we refer to as the ``dispersive-interaction'' frame), after the rotating wave approximation the full system Hamiltonian is then
\be
\hat{H}_{\rm I}(t) = \eps e^{-i\delta t} e^{i\chi\hat{\sigma}_zt}\hat{a}+\eps^*e^{i\delta t}e^{-i\chi\hat{\sigma}_zt}\hat{a}^{\dagger},
\ee
where $\delta = \ww_{\rm c} - \ww_{\rm d}$ is the cavity-drive detuning, and the subscript ``I'' labels the interaction picture. In the dispersive-interaction frame, after a time $T$ the state of the full system is described by
\begin{align}
\nonumber \ket{\psi_{\rm D}' (T)} &= \hat{U}_{\rm I}(T,0)\ket{\psi' (0)}\\
&=\mathcal{T}\exp\left\{-i\int_0^T \hat{H}_{\rm I}(t)\right\}\ket{\psi' (0)},
\end{align}
where $\mathcal{T}$ is the usual time ordering operator. 

To calculate $\hat{U}_{\rm I}(T,0)$ exactly we will use the Magnus expansion \cite{Magnus:1954yu}, given by
\be
\hat{U}_{\rm I}(T,0) = \exp\left(\sum_{k=1}^{\infty}\Omega_{k}(T,0)\right)
\ee
where $\Omega_k(T,0)$ is the $k$'th order Magnus generator. For our system these generators are zero for $k>2$, and so $\hat{U}_{\rm I}(T,0)$ can be calculated exactly. The first order Magnus generator is given by an expression proportional to the average Hamiltonian
\begin{align}
\nonumber\Omega_1(T,0) &= -i\int_0^T {\rm d}t \  \hat{H}_{\rm I}(t) = \ket{g}\bra{g}(\alpha_{\rm g}(T)\hat{a}^\dagger-\alpha^*_{\rm g}(T)\hat{a}) \\&+\ket{e}\bra{e} (\alpha_{\rm e}(T)\hat{a}^\dagger-\alpha^*_{\rm e}(T)\hat{a}),
\end{align}
where
\begin{align}
\nonumber &\al_{\rm g}(T) = \frac{-\eps^*\left(e^{i\left(\delta-\chi\right)T}-1\right)}{\delta-\chi}, \\ &\al_{\rm e}(T) = \frac{-\eps^*\left(e^{i\left(\delta+\chi\right)T}-1\right)}{\delta+\chi}.
\label{eqn:Alphas} 
\end{align}
As $[\hat{H}_{\rm I}(t_1),\hat{H}_{\rm I}(t_2)] \propto f(\sigma_z) \otimes \mathbb{I}$ which commutes with $H_{\rm I}(t_3)$ the Magnus expansion truncates at second order, and the second order generator results in a qubit-state dependent relative Stark phase. Also, as $\Omega_2(T,0)$ commutes with $\Omega_1(T,0)$ the full evolution operator is
\begin{align}
\nonumber&\hat{U}_{\rm I}(T,0) = \exp\left\{\Omega_1(T,0)\right\}\exp\left\{\Omega_2(T,0)\right\}\\&= \left(\ket{g}\bra{g} \hat{D}(\al_{\rm g}(T)) + \ket{e}\bra{e}\hat{D}(\al_{\rm e}(T))\right)  e^{iF(\sigma_zT)} \label{eqn:Udrive}
\end{align}
where $\hat{D}(\beta) = \exp\left\{\beta\hat{a}^\dagger - \beta^*\hat{a}\right\}$ is the usual displacement operator, and $  e^{iF(\sigma_zT)}$ is the phase from the $\Omega_2(T,0)$ term, shown explicitly in equation (\ref{eqn:A2}).

In the lab frame of equation (\ref{eqn:JC}) we consider an initial state given by $\ket{g,0}$, which is the ground state of the Jaynes-Cummings Hamiltonian. This state is a dark state and therefore invariant under both the transformation into the dispersive frame $\hat{U}_{\rm D}(t)$ and the transformation into the interaction frame $\exp\left\{i\hat{H}_{\rm D}t\right\}$, and so in the dispersive-interaction frame the initial state is $\ket{\psi'(0)} = \ket{g,0}$ (we work in the lab frame basis throughout). 

\begin{figure}
\includegraphics[width=\columnwidth]{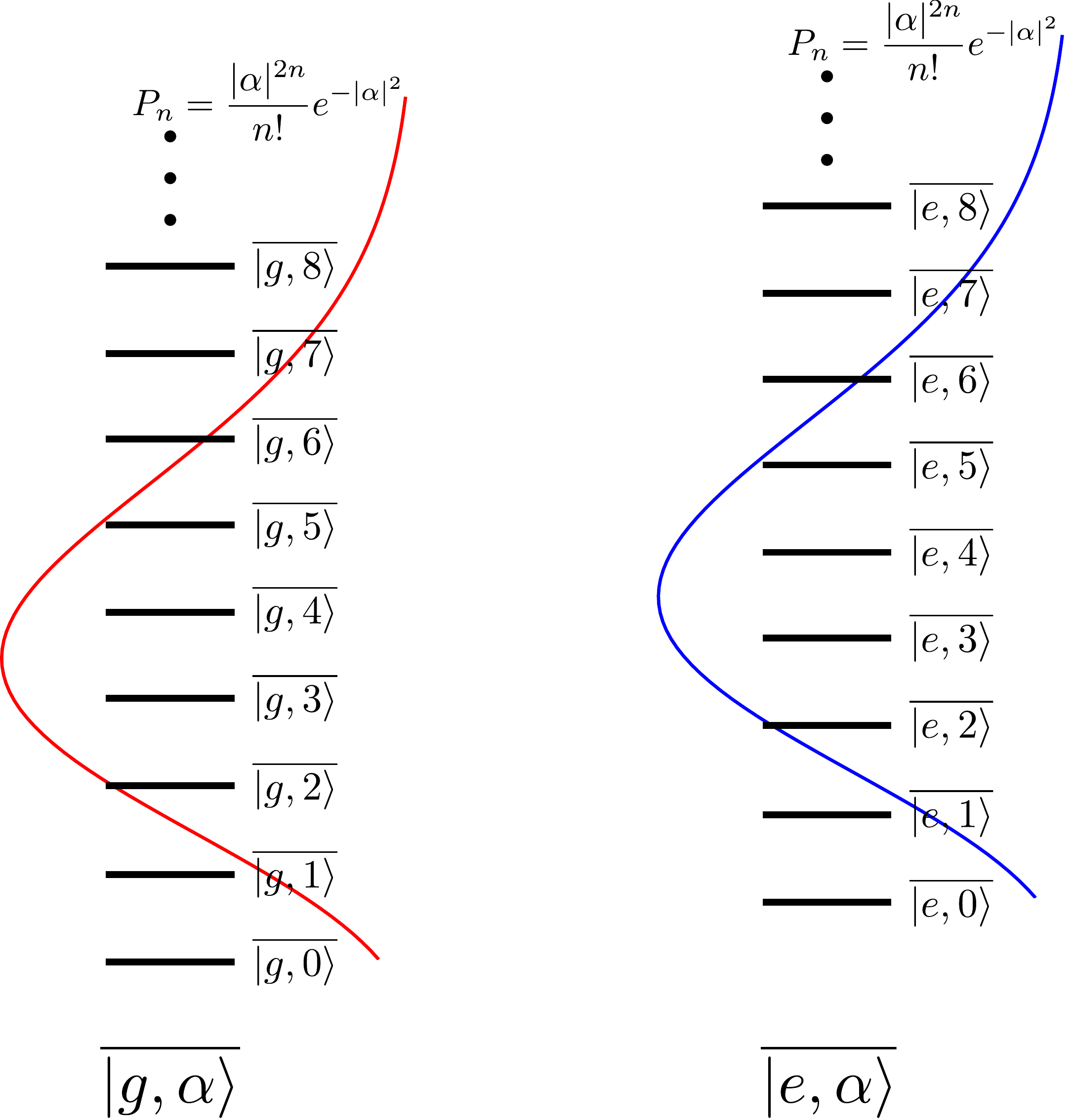}
\caption{Schematic diagrams of the dressed coherent states $\overline{\ket{g,\al}}$ and $\overline{\ket{e,\al}}$ for $\abs{\al}^2 =4$. The red and blue curves are the Poisson distribution of superposition coefficients, $P_n = \abs{\overline{\braket{g/e,n}{g/e,\al}}}^2$.}
\label{fig:DCdiagram}
\end{figure}

We can then calculate the final state in the dispersive-interaction frame
\begin{align}
\nonumber\ket{\psi_{\rm D}' (T)} &= \left(\ket{g}\bra{g}\hat{D}(\al_{\rm g}(T)) +\ket{e}\bra{e} \hat{D}(\al_{\rm e}(T))\right)\ket{g,0}\\&= \ket{g, \al_{\rm g}(T)},
\end{align}
where as our initial state is not a qubit superposition state the effect of the second order Magnus term is a global phase that can be ignored. Next we transform this trivially out of the interaction frame back to the dispersive frame
\begin{align}
\nonumber\ket{\psi_{\rm D} (T)} &= \exp\left\{-i\hat{H}_{\rm D}T\right\}\ket{\psi_{\rm D}' (T)} \\ \nonumber&= \ket{g}e^{-\abs{\al_{\rm g}(T)}^2}\sum_n \frac{\al_{\rm g}^n(T)}{\sqrt{n!}} e^{-i\left( \ww_{\rm c} -\chi\right)\hat{a}^{\dagger}\hat{a}T}\ket{n} \\
&=\ket{g,\al_{\rm g}(T)e^{-i(\ww_{\rm c}-\chi)T}} = \ket{g,\tilde{\al}_{\rm g}(T)}. 
\end{align}
To return to the lab frame we must apply the inverse dispersive transformation $\hat{U}_{\rm D}^\dagger$, and using equation (\ref{eqn:gUDdag}) we see that the final state in the lab frame is
\begin{align}
\nonumber\ket{\psi (T)} &= \hat{U}^{\dagger}_{\rm D}\ket{g,\tilde{\al}_{\rm g}(T)} = e^{-\abs{\tilde{\al}_{\rm g}(T)}^2}\sum_n \frac{\tilde{\al}_{\rm g}^n(T)}{\sqrt{n!}} \hat{U}^{\dagger}_{\rm D}\ket{g,n} \\&= e^{-\abs{\tilde{\al}_{\rm g}(T)}^2}\sum_n \frac{\tilde{\al}_{\rm g}^n(T)}{\sqrt{n!}}\overline{\ket{g,n}} \label{eqn:apDC}.
\end{align}
The qubit-cavity state of equation (\ref{eqn:apDC}) is the dressed coherent state $\overline{\ket{g,\tilde{\al}_{\rm g}(T)}}$, where in general a dressed coherent state has the form
\begin{align}
\overline{\ket{g/e,\al}} = e^{-\frac{\abs{\al}^2}{2}}\sum_{n}\frac{\al^n}{\sqrt{n!}}\overline{\ket{g/e,n}},
\label{eqn:GenericDC}
\end{align}
where $\overline{\ket{g/e,n}}$ is given exactly be equation (\ref{eqn:Geig})/(\ref{eqn:Eeig}), and to first order in $\lambda$ by equation (\ref{eqn:aGeig})/(\ref{eqn:aEeig}). To the best of our knowledge this state was first described in Ref.~\cite{Sete:2013rt}. Schematic diagrams of the dressed coherent states $\overline{\ket{g,\al}}$ and $\overline{\ket{e,\al}}$ for $\abs{\al}^2 = 4$ are shown in FIG.~\ref{fig:DCdiagram}. The curves in FIG.~\ref{fig:DCdiagram} represent the Poissonian weights of the sum in equation (\ref{eqn:GenericDC}), and highlight the fact that the dressed coherent state has the same distribution of superposition coefficients as the coherent state, only for the dressed states instead of the bare states.

As we have just shown, if the system starts in the state $\ket{g,0}$ then the final state of the system after a classical cavity drive of length $T$ will be the dressed coherent state $\overline{\ket{g,\tilde{\al}_{\rm g}(T)}}$. Similarly, if the system starts in its first excited state, given by $\overline{\ket{e,0}}$ in the lab frame, then the final state after classical drive will be the state $\overline{\ket{e,\tilde{\al}_{\rm e}(T)}}$ (see appendix \ref{app:Excited} for further details). We will now discuss an intuitive physical understanding of the dressed coherent state, and compare the analytic results to full numerical simulations.

\subsection{Effective Qubit Drive}
To understand the dressed coherent state for $\lambda \ll 1$ it is useful to expand the state of equation (\ref{eqn:GenericDC}) in powers of $\lambda$ to obtain
\begin{align}
\nonumber&\overline{\ket{g,\beta}} = e^{-\frac{\abs{\beta}^2}{2}}\sum_{n}\frac{\beta^n}{\sqrt{n!}}\Bigg[\left(1-\lambda^2\frac{n}{2}\right)\ket{g,n}
\\ \nonumber&-\lambda\sqrt{n}\ket{e,n-1}\Bigg] + \mathcal{O}\left(\lambda^3\right)
\\ \nonumber&=  \left(\ket{g}-\lambda\beta\ket{e}\right)\ket{\beta} \\ &- \frac{\lambda^2}{2}e^{-\frac{\abs{\beta}^2}{2}}\sum_{n}\frac{\beta^n}{\sqrt{(n-1)!}}\sqrt{n}\ket{g,n} + \mathcal{O}\left(\lambda^3\right), \label{eqn:apDg}
\end{align}
and we see that to lowest nontrivial order in $\lambda$ the qubit is effectively in the superposition state $\left(\ket{g} -\lambda\beta\ket{e}\right)/\sqrt{\mathcal{N}}$, with the normalization factor $\mathcal{N} = 1 +\lambda^2\abs{\beta}^2$. Thus, to lowest order in $\lambda$, when the qubit-cavity system is driven by a classical cavity drive (creating a dressed coherent state) the qubit is effectively weakly driven on resonance via the Hamiltonian (in the interaction picture)
\be
\hat{H}_{\rm Eff} = q_0\hat{\sigma}^{-} + q^*_0\hat{\sigma}^{+},
\label{eqn:EffQdrive}
\ee
for a time $T$ as before, where the effective qubit drive strength is given by $q_0 = (i\beta^*\lambda)/T$ for the dressed coherent state $\overline{\ket{g,\beta}}$.

It is important to note that under the dispersive approximation the cavity drive of equation (\ref{eqn:CavDri}) is only the leading order term, and that the next order correction in the dispersive frame, of order $\lambda$, is a qubit drive. However, this drive is at a frequency $\ww_{\rm d}$, which will not be resonant with the qubit transition frequency, and as a result this qubit drive has no net effect on relevant timescales. For this reason we have not included this off-resonant qubit drive in our analtyical calculations. 

The effective resonant qubit drive described here by equation (\ref{eqn:EffQdrive}) is not due to this first order term in the dispersive approximation of the cavity drive Hamiltonian, as it occurs even when only the zeroth order term of the dispersive frame cavity drive is considered, as in equation (\ref{eqn:CavDri}). It is uniquely an effect of considering the lab frame state for qubit-cavity interaction with a driven cavity, and results from interactions between the cavity and the qubit.

\subsection{Numerical Simulations}
\label{sec:DressedNums}

In section \ref{sec:Cdrive} we have shown that under the dispersive approximation done completely, the final state in the lab frame of a qubit-cavity system after a classical cavity drive is a dressed coherent state. While for $\lambda \ll 1$ and $\abs{\al}^2 \ll n_{\rm crit}$ it is sufficient to keep terms up to order $\lambda^2$ and obtain the Hamiltonian of equation (\ref{eqn:Disp}), it is worthwhile to examine what effect the neglected higher order terms have on the final state. To do so, we numerically investigate time evolution induced by the Hamiltonian
\begin{align}
\nonumber\hat{H}_{\rm T}(t) &= \ww_{\rm c}\hat{a}^{\dagger}\hat{a} - \frac{\ww_{\rm q}}{2}\hat{\sigma}_z + g\left(\hat{\sigma}^{-}\hat{a}^{\dagger}+\hat{\sigma}^{+}\hat{a}\right) \\&+ \left(\eps e^{i\ww_{\rm d} t}\hat{a}+\eps^*e^{-i\ww_{\rm d} t}\hat{a}^{\dagger}\right)\Theta(t-T), \label{eqn:TotalHam}
\end{align}
for the lab frame initial states $\ket{g,0}$ and $\overline{\ket{e,0}}$, with $\ww_{\rm d} = \ww_{\rm c} - \chi$ and $\ww_{\rm d} = \ww_{\rm c} + \chi$ respectively. We simulate over a range of $\lambda$, $\eps$, and target $\tilde{\al}_{\rm g/e}(T)$ to see how these parameters affect the accuracy of the dressed coherent state.

We compare the numerically created states with the dressed coherent states $\overline{\ket{g,\tilde{\al}_{\rm g}(T)}}$ and $\overline{\ket{e,\tilde{\al}_{\rm e}(T)}}$, as well as the undressed product states  $\ket{g,\tilde{\al}_{\rm g}(T)}$ and $\ket{e,\tilde{\al}_{\rm e}(T)}$ (all in the lab frame), by calculating the overlaps
\begin{align}
&\mathcal{F}_{\rm D}^{g/e}(\abs{\al}^2,\eps,\lambda) = \Big|\bra{\psi(T)}\overline{\ket{g/e,\tilde{\al}_{\rm g/e}(T)}}\Big|^2, \label{eqn:FidD} \\
&\mathcal{F}^{g/e}(\abs{\al}^2,\eps,\lambda) = \Big|\bra{\psi(T)}\ket{g/e,\tilde{\al}_{\rm g/e}(T)}\Big|^2, \label{eqn:FidU}
\end{align}
where $\ket{\psi(T)}$ is the state created by numerical simulation, and we have set that for either initial state the target coherent state amplitude is the same, i.e. $\abs{\tilde{\al}_{\rm g}(T)} = \abs{\tilde{\al}_{\rm e}(T)} = \abs{\al(T)}$. The phase of $\al(T)$ has no impact on the fidelity and so the fidelity depends only on $\abs{\al}^2$.

\begin{figure*}
\subfigure{
\includegraphics[width =\columnwidth]{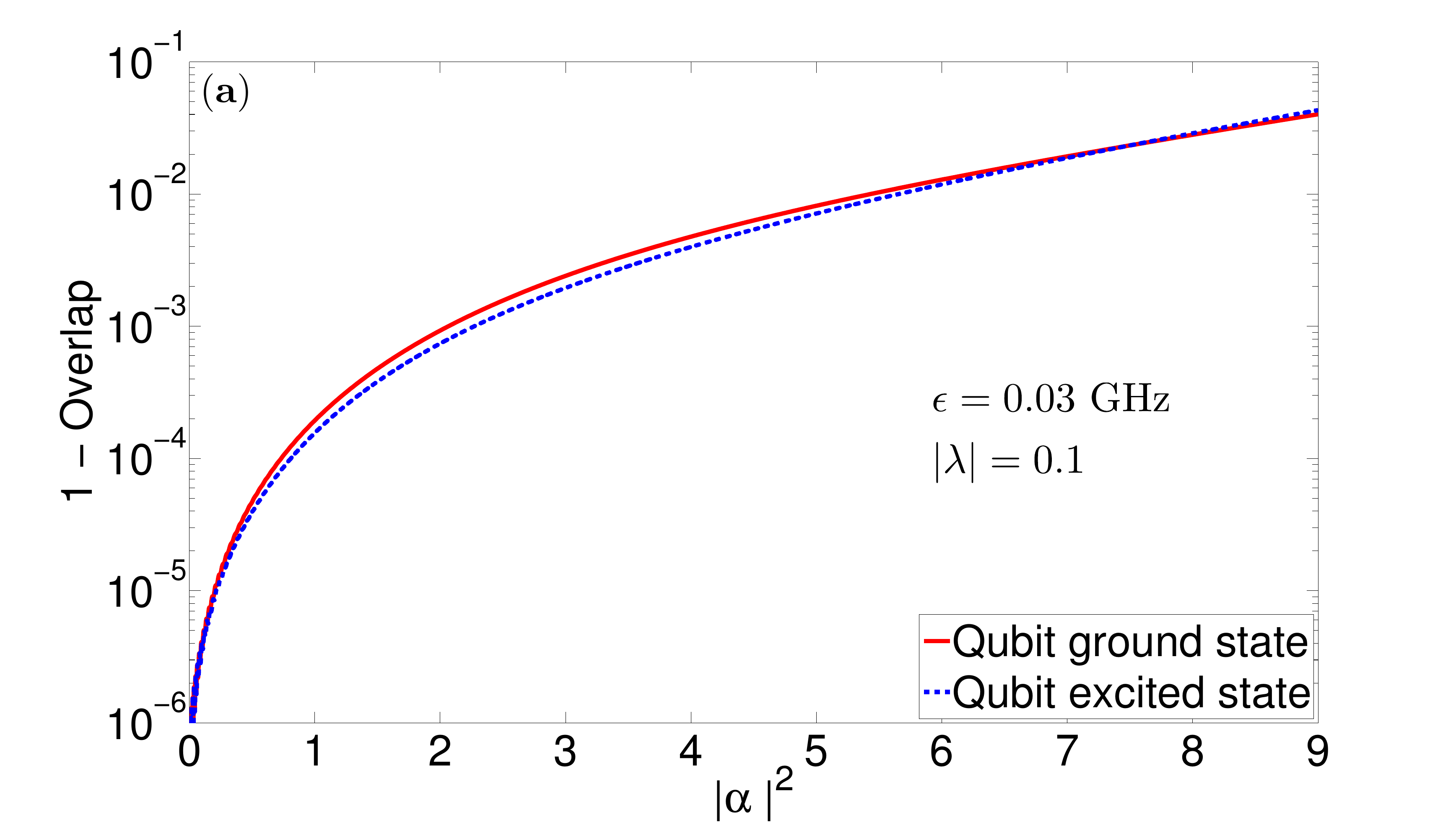}
\label{fig:PowerO}}
\subfigure{
\includegraphics[width =\columnwidth]{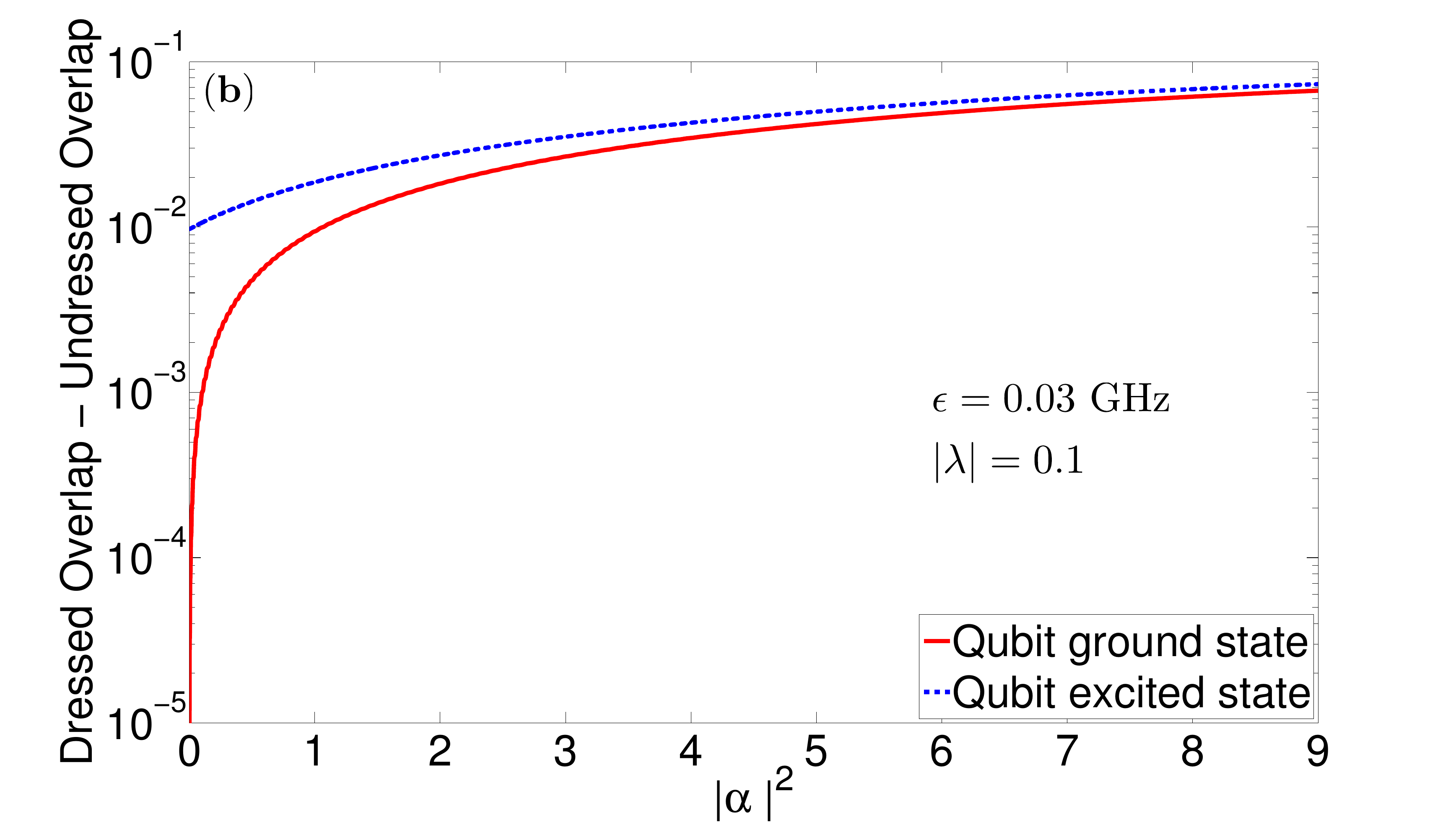}
\label{fig:PowerOD}}
\subfigure{
\includegraphics[width =\columnwidth]{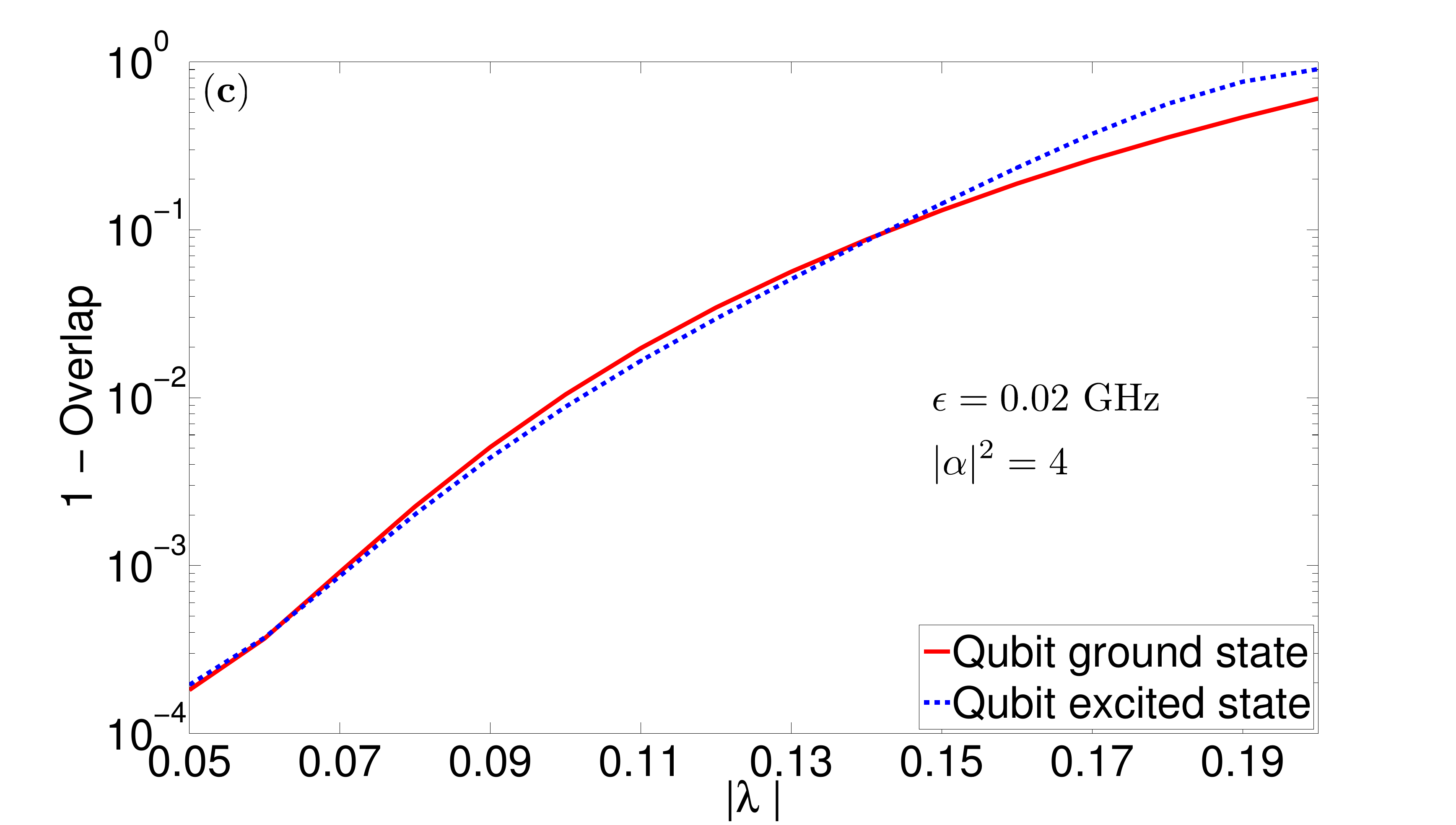}
\label{fig:GammaO}}
\subfigure{
\includegraphics[width =\columnwidth]{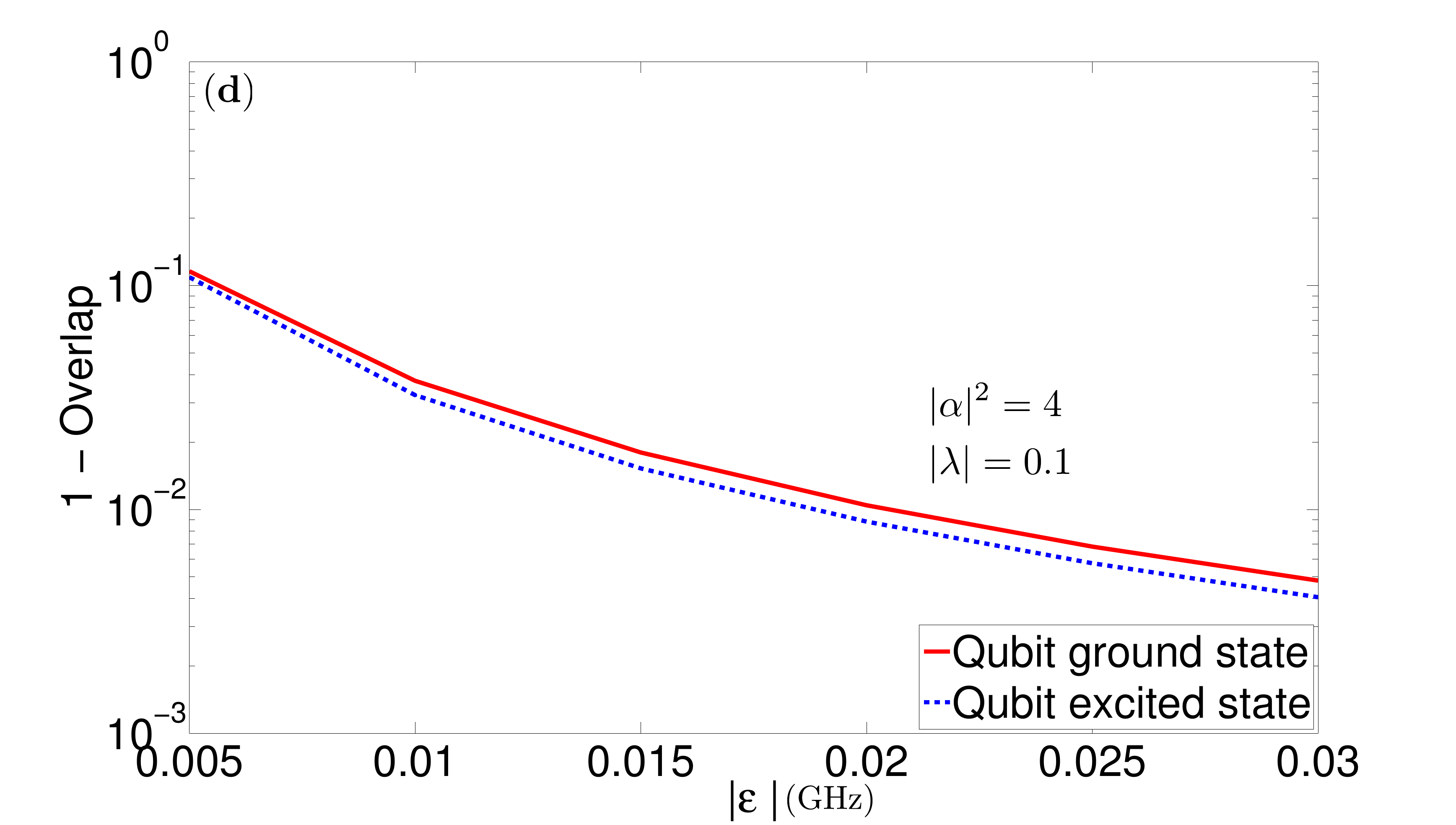}
\label{fig:DriveO}}
\caption{Comparison of the numerical state created by simulation of equation (\ref{eqn:TotalHam}) with the dressed coherent state as a function of {\bf (a)} photon number $\abs{\al}^2$, with $\eps$ and $\lambda$ constant, {\bf (c)} $\lambda$, with $\eps$ and $\abs{\al}^2$ constant, and {\bf (d)} drive strength $\eps$, with $\abs{\al}^2$ and $\lambda$ constant. {\bf (b)} The difference in overlap with the numerical state between the dressed coherent state and the undressed product state as a function of photon number $\abs{\al}^2$, with $\eps$ and $\lambda$ constant. Values of $\abs{\al}^2$, $\lambda$, and $\eps$ chosen to be commensurate with current experiments. Additional phase matching was required to obtain high fidelity (see appendix \ref{app:PhaseMatch}).}
\label{fig:Overlaps}
\end{figure*}

Figures \ref{fig:PowerO}, \ref{fig:GammaO}, and \ref{fig:DriveO} show $1 - \mathcal{F}_{\rm D}^{g/e}(\abs{\al}^2,\eps,\lambda)$ for varying $\abs{\al}^2$, $\lambda$, and $\eps$ respectively, with the non-varying parameters held at the constant values indicted on the figures. Fidelities for both the ground or excited dressed coherent state are plotted. Due to decreasing validity of the dispersive approximation, both FIGs.~\ref{fig:PowerO} and \ref{fig:GammaO} show a decreasing overlap as either $\abs{\al}^2$ or $\lambda$ increase. As is to be expected, increasing $\lambda$ is more detrimental to the agreement between the numerical state and the dressed coherent state, as a larger $\lambda$ requires fewer photons in the cavity for terms beyond order $\lambda^2$ in the full Hamiltonian to become relevant. For increasing $\abs{\al}^2$, FIG.~\ref{fig:PowerO} shows that even for a high average photon number of $\abs{\al}^2 = 9$ the numerical state still has an overlap greater than 90\% with a dressed coherent state. 

Interestingly, FIG.~\ref{fig:DriveO} shows that for increasing drive strength $\eps$, the overlap between the dressed coherent state and the numerical state actually increases. This can likely be understood by competition between the cavity drive and higher order effects beyond the dispersive approximation, which will both be off-diagonal in the basis of equation (\ref{eqn:Disp}). The stronger the drive, the more it dominates this competition, which therefore diminishes the effect of the higher order correction terms, leading to a state closer to the dressed coherent state.

Finally, FIG.~\ref{fig:PowerOD} shows the difference in fidelity between the dressed coherent state and the undressed product state, i.e. $\mathcal{F}_{\rm D}^{g/e}(\abs{\al}^2,\eps,\lambda) - \mathcal{F}^{g/e}(\abs{\al}^2,\eps,\lambda)$, as a function of $\abs{\al}^2$. As can be seen, the difference is always positive, and the dressed coherent state is always a better description of the numerical state than the undressed product state.

\section{Implications and Applications}
\label{sec:discuss}

\subsection{Qubit Operations}
\label{sec:Qdrive}

In this section we examine a simple case of qubit operations performed on a dressed coherent state to emphasize the dependence of these operations on both the amplitude and phase of the cavity state. Such interactions could occur in a set-up that involved parallel operations on a qubit and a cavity used as a bus \cite{Galiautdinov:2012nr} or as a quantum memory \cite{Leghtas:2013kq}. We consider a primitive for qubit operations that, while itself not necessarily useful or interesting, elucidates the relevant physics involved, such that useful and interesting applications may be developed from it in future work. Starting in the dressed coherent state $\overline{\ket{g,\beta}}$ (created as described in section \ref{sec:Cdrive}), we attempt to rotate the reduced state of the qubit to as close to the excited state as possible. We will show that the excitation probability will depend on both the amplitude and phase of $\beta$.

We begin by expanding the inital dressed coherent state $\overline{\ket{g,\beta}}$ to lowest order in $\lambda$, as in equation (\ref{eqn:apDg}). In this case, the reduced state of the qubit is a weakly rotated qubit state, with the phase of this rotation proportional to the phase of $\beta$. For a weakly rotated qubit state, the probability amplitude upon further rotating the qubit depends on the phase of the applied drive relative to the phase of the initial qubit rotation, and therefore, the success of rotating to the excited state from the state $\overline{\ket{g,\beta}}$ will depend on the phase of $\beta$. For purely real (imaginary) $\beta$, the reduced qubit state lies in the $x$-$z$ ($y$-$z$) plane on the Bloch sphere, and rotation about the $x$ ($y$) axis cannot transform the qubit state to the state $\ket{e}$, while rotation about the $y$ ($x$) axis can. For general $\beta$ the superposition is along the plane defined by the $z$ axis and the line $\{(x,y) = (\cos(\phi_{\beta})s, \sin(\phi_{\beta})s),\ \forall s \in \mathbb{R} \}$ in the $x$-$y$ plane, which depends on the phase of $\beta$, given by $\phi_{\beta}$. In this case perfect state transfer to $\ket{e}$ is not possible by rotation along any axis other than that defined by $\phi_{\beta}+\pi/2$, given by the line $\{(x,y) = (\sin(\phi_{\beta})s, \cos(\phi_{\beta})s),\ \forall s \in \mathbb{R} \}$.

More rigorously, we consider the qubit-drive Hamiltonian to first order in the dispersive frame given by
\be
\hat{H}_{\rm Q} = \eta e^{-i\ww t}\hat{\sigma}^{+} + \eta^*e^{i\ww t}\hat{\sigma}^{-}.
\label{eqn:ApQdrive}
\ee
In the interaction picture with respect to the dispersive Hamiltonian of equation ($\ref{eqn:Disp}$), this Hamiltonian becomes
\be
\hat{H}'_{\rm Q} = \eta e^{i\nu t} e^{i2\chi\hat{n}t}\hat{\sigma}^{+} + \eta^*e^{-i\nu t}e^{-i2\chi\hat{n}t}\hat{\sigma}^{-},
\ee
where $\hat{n} = \hat{a}^{\dagger}\hat{a}$ and $\nu = \ww_{\rm q} + \chi - \ww$. To calculate the evolution operator, we follow the same procedure as in section \ref{sec:Cdrive} using the Magnus expansion. Unfortunately, the Magnus generators $\Omega_{n}$ do not vanish for any finite order $n$, due to the nonlinearity of the qubit. To account for this we choose our evolution time to be short enough (roughly one period for the drive frequency $\ww$) that Magnus terms beyond first order have minimal contribution to the evolution, as the magnitude of their effects is small and only becomes relevant after accumulating for a long period of time. With this in mind, the analytical results presented below are meant to be instructive and to highlight the important physical effects, rather than to have high accuracy.

The result of this calculation (see appendix \ref{app:DriveQ} for further details) is a very complicated expression for the excited state probability as a function of time, given by equation (\ref{eqn:Eprob}). In order to gain some intuitive understanding of the rigorous result, we set $\ww=\ww_{\rm q}$ and go to the limit where $\chi,\lambda \rightarrow 0$, in which case, the probability of finding the qubit in the excited state at time $\tau$ is given by
\begin{align}
\nonumber{\rm P}_{e}(\tau) &= \left(1-\lambda^2\right)\sin^2\left(\eta\tau\right) + \lambda^2\abs{\beta}^2\cos(2\eta\tau) \\\nonumber&+ \lambda\sin(2\eta\tau)\Big({\rm Im}(\beta e^{-i\varphi})\cos(\Delta \tau)\\&+{\rm Re}(\beta e^{-i\varphi})\sin(\Delta\tau)\Big),
\label{eqn:EprobAp} 
\end{align}
where $e^{i\varphi}=\eta/\abs{\eta}$, and $\Delta = \ww_{\rm q}-\ww_{\rm c}$ as before. From the last term in equation (\ref{eqn:EprobAp}) we see that the excited state probability depends not only on the photon number in the cavity, but also the interplay between the coherent state phase and the phase of the drive, through the unequal dependence on ${\rm Re}(\beta e^{-i\varphi})$ and ${\rm Im}(\beta e^{-i\varphi})$. This effect can be understood as interference between the effective qubit drive of equation (\ref{eqn:EffQdrive}) caused by the dressed coherent state and the applied qubit drive of equation (\ref{eqn:ApQdrive}), in a manner similar to coherent destruction of tunneling \cite{Grossmann:1991jk}.

\begin{figure}
\includegraphics[width=\columnwidth]{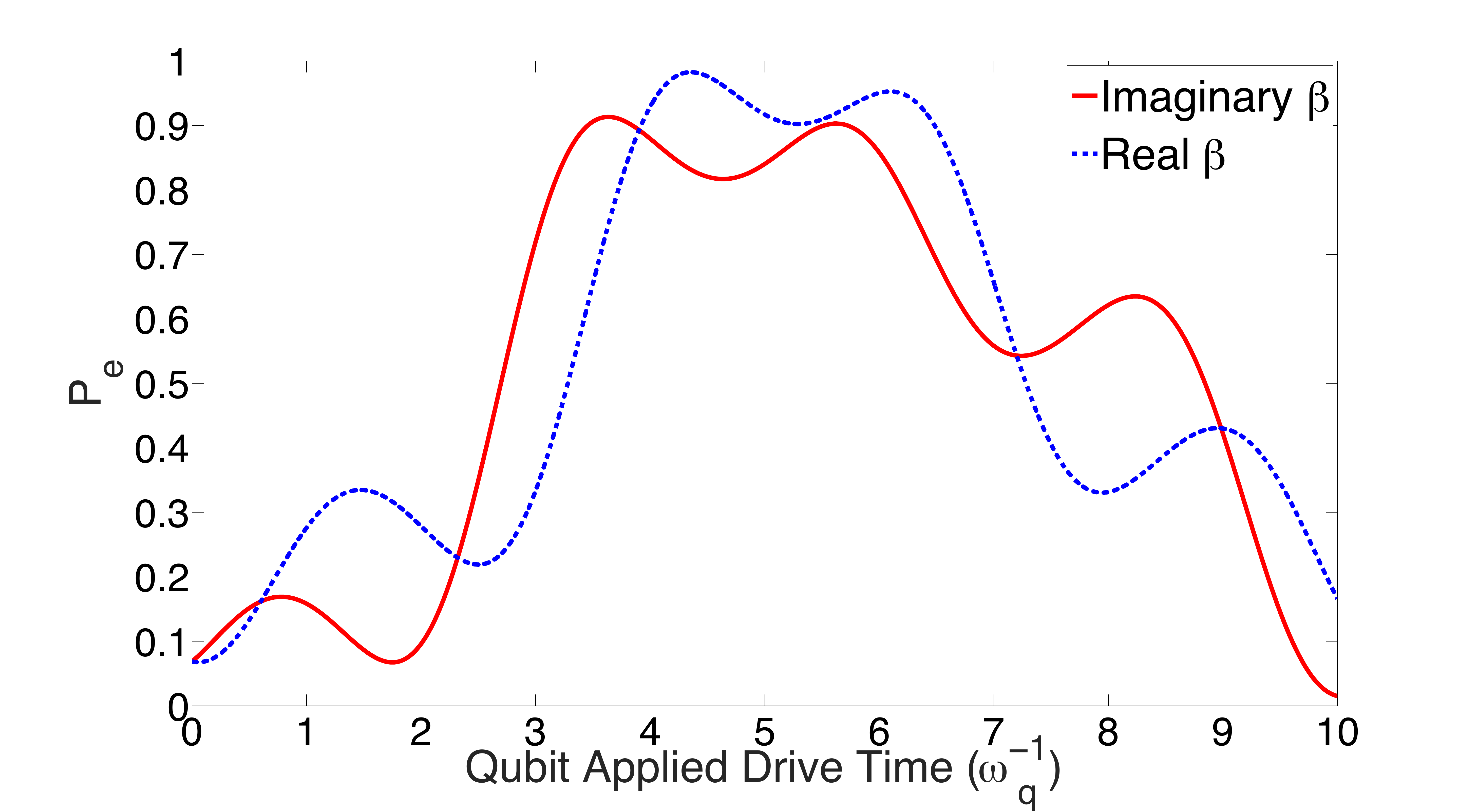}
\caption{Qubit excited state probability as a function of the time of the applied qubit drive, starting in a dressed coherent state with either purely real or purely imaginary $\beta$ for $\abs{\beta}^2=4$. The drive strength is strong, $\abs{\eta} = 0.05\ww_{\rm q}$, so that phase effects are clearly visible. The drive frequency is set to the average shifted qubit frequency, $\ww = \ww_{\rm q} + 2\left(\abs{\beta}^2+1\right)$, $\eta$ is purely real (i.e. $\abs{\eta} = \eta$), and $\lambda = 0.1$ as elsewhere.}
\label{fig:QExProb}
\end{figure}

Unfortunately, while easy to understand, equation (\ref{eqn:EprobAp}) is not very accurate in the relevant parameter regimes (due to the approximations made), and the error in truncating the Magnus expansion at first order grows for longer times. However, the analytical calculations leading to equation (\ref{eqn:EprobAp}) were done to distill the relevant physical effects and present them in an understandable manner, not to obtain highly accurate results. To accurately test the phase dependence effects, we numerically simulate a qubit-cavity system with classical qubit drive (without making the dispersive approximation). As an initial state, we start with the state created by simulation of equation (\ref{eqn:TotalHam}) of section {\ref{sec:DressedNums}}, which has high overlap with a dressed coherent state. Figure \ref{fig:QExProb} shows the excited state probability as a function of time, starting in a dressed coherent state with either purely real or imaginary $\beta$, and with a purely real qubit drive $\eta$. 

In addition to photon number effects, where the qubit resonance frequency is modified by the photon number in the cavity \cite{Schuster:2007qf}, FIG. \ref{fig:QExProb} shows that the excited state probability's time evolution depends also on the phase of the dressed coherent state amplitude $\beta$. This agrees with the intuitive conclusions drawn previously, using the approximation of equation (\ref{eqn:apDg}), where one considers the dressed coherent state to be a coherent state in the cavity, and a weakly rotated qubit.

\subsection{Dispersive Multi-Qubit Readout with a Threshold Detector - Limited Contrast}

In recent proposals for dispersive single qubit readout \cite{Govia:2014fk}, and qubit parity readout \cite{Govia:2015fk}, protocols were developed that conditionally populate the cavity dependent on the state (parity) of the qubit(s), after which, by using a threshold photon counter to distinguish between the qubit dependent cavity states, the state (parity) of the qubit(s) can be non-destructively measured. A subsequent coherent cavity drive of opposite phase removes the cavity occupation. Such conditional cavity occupation can be achieved by setting $\ww_{\rm d} = \ww_{\rm c} - \chi$ in equation (\ref{eqn:Alphas}), and the correct choice of $T$ such that $\tilde{\al}_{\rm g}(T) \neq 0$ while  $\tilde{\al}_{\rm e}(T) = 0$. 

For single-qubit readout, starting from an initial state in the lab frame of either $\ket{g,0}$ or $\ket{e,0}$ and using equations (\ref{eqn:apDC}) and (\ref{eqn:eDress}), after classical cavity drive the qubit state dependent qubit-cavity states are
\begin{flalign}
&\ket{\psi(T)} = \overline{\ket{g,\tilde{\al}_{g}(T)}}, \\
&\ket{\Psi (T)} = \cos\left(\lambda\right)\overline{\ket{e,0}} - e^{i(\ww_{\rm q}+ \chi)}\sin\left(\lambda\right)\hat{U}^{\dagger}_{\rm D}\ket{g}\ket{\xi(T)}, \label{eqn:parityE}
\end{flalign}
for the suitable choice of $T$ that ensures $\tilde{\al}_{\rm e}(T) = 0$. Misidentification of the qubit excited state as the ground state occurs when there is spurious photon population in the cavity when the qubit is in its excited state. As can be calculated using equation (\ref{eqn:parityE}) this spurious photon population is given by
\begin{align}
\nonumber N &= \bra{\Psi(T)}\hat{a}^\dagger\hat{a}\ket{\Psi (T)} 
\approx \sin^2\left(\lambda\right)\cos^2\left(\lambda\right)\bra{1}\hat{a}^\dagger\hat{a}\ket{1} \\ \nonumber&+ \sin^2\left(\lambda\right)\bra{g}\bra{\xi(T)}\hat{U}_{\rm D}\hat{a}^\dagger\hat{a}\hat{U}^{\dagger}_{\rm D}\ket{g}\ket{\xi(T)}\\
\nonumber&\approx \sin^2\left(\lambda\right)\left(\cos^2\left(\lambda\right) +\bra{\xi(T)}\hat{a}^\dagger\hat{a}\ket{\xi(T)} \right) \\
&= \sin^2\left(\lambda\right)(\cos^2\left(\lambda\right) +1 + \abs{\al_{g}}^2)
\end{align}
where in the first approximation we have ignored the cross terms as the photon occupation of $\ket{\xi(T)}$ is much greater than $1$, and in the second approximation we have assumed the dispersive transformation only slightly modifies the average photon number of $\ket{\xi(T)}$. The photon occupation of $\ket{\xi(T)}$ is high since it is the single photon Fock state displaced by $\hat{D}(\al_{g}(T))$, and $\al_{g}(T) > 0$ for this protocol. Extending this effect to multiple qubits in their excited states explains the even parity misidentification error for four qubit parity measurement seen in Ref.~\cite{Govia:2015fk}, as it is the four-qubit generalization of the state $\ket{\xi(T)}$ that leads to spurious detections when the qubits are in an even parity state. The exact value of this error is dependent on the nature of the threshold photon counter used \cite{Govia2012,Govia2014}.

\section{Conclusion}
\label{sec:conc}

In the work presented here, we have undertaken a critical and careful examination of the dispersive limit, dispersive approximation, and the dispersive Hamiltonian, in order to understand the distinction and commonalities between these commonly used terms. We have found that one obtains the dispersive Hamiltonian after making the dispersive approximation, which requires both being in the dispersive limit of the Jaynes-Cummings interaction, and a frame transformation to the dispersive frame. The dispersive Hamiltonian therefore describes evolution of the system in the dispersive frame, and in order to obtain results valid in the lab frame, one must apply the inverse frame transformation.

This has a profound effect on the description of the system state, and in particular, we have found that after a classical drive is applied to the cavity, the state of the system in the lab frame is accurately described by the dressed coherent state $\overline{\ket{g/e,\al}}$, not, as often used, the product state $\ket{g/e}\ket{\al}$. The entanglement present in the dressed coherent state will affect all future operations on the qubit and the cavity. We have shown how this is relevant to rotations of the qubit state, in particular that the probability of rotating the qubit from its ground to excited state (or vice versa) depends on the phase difference between $\al$ and the applied qubit drive. We have also explained the limit in measurement contrast for readout of the qubit state via a cavity and a threshold photon counter, reported in Ref.~\cite{Govia:2015fk}. Future work will continue to explore the effects of the dressed coherent state on contemporary quantum information protocols.

\section*{Acknowledgments}

The authors acknowledge insightful discussions with Bruno G. Taketani. Supported by the Army Research Office under contract W911NF-14-1-0080 and the European Union through ScaleQIT. LCGG acknowledges support from NSERC through an NSERC PGS-D.

\appendix
\section{Dressed Coherent State for An Excited Qubit}
\label{app:Excited}
In this appendix we consider starting in the initial state in the lab frame of the bare excited qubit state $\ket{e,0}$, which can be prepared by initialization to the state $\ket{g,0}$ followed by a fast, nonadiabatic pulse on the qubit. In the dispersive frame the initial state will be $\ket{\Psi_{\rm D}(0)} = U_{\rm D}\ket{e,0} = \cos\left(\lambda\right)\ket{e,0} - \sin\left(\lambda\right)\ket{g,1}$. Transforming into the interaction frame, the state is unchanged since $t=0$. To calculate the final state, we will need the second order Magnus generator, given by
\begin{align}
\nonumber\Omega_2(T,0) &= -\frac{1}{2}\int_0^T {\rm d}t_1 \int_0^{t_1} {\rm d}t_2 \  \left[\hat{H}_{\rm I}(t_1),\hat{H}_{\rm I}(t_2)\right] \\&= i\frac{\abs{\eps}^2}{\chi^2}\Big(\sin(\chi T\sigma_z)-\chi T\sigma_z\Big)
\end{align}
From this we can calculate the qubit-state dependent phase due to the second order Magnus generator, given by
\be
 e^{iF(\sigma_zT)} = \exp\left(i\frac{\abs{\eps}^2}{\chi^2}\Big(\sin(\chi T\sigma_z)-\chi T\sigma_z\Big)\right),
\label{eqn:A2}
\ee
and make the identification
\be
F(\sigma_zT) = \frac{\abs{\eps}^2}{\chi^2}\Big(\sin(\chi T\sigma_z)-\chi T\sigma_z\Big).
\ee

Now using equation (\ref{eqn:Udrive}), at $t = T$ the system state in the dispersive-interaction frame is given by
\begin{align}
\nonumber\ket{\Psi_{\rm D}'(T)} &= \cos\left(\lambda\right)\ket{e,\al_{\rm e}(T)} \\&- e^{i[F(T)-F(-T)]}\sin\left(\lambda\right)\ket{g}\hat{D}(\al_{\rm g}(T))\ket{1},
\end{align}
where we have factored out a global phase. Transforming back into the dispersive frame we have:
\begin{align}
\nonumber&\ket{\Psi_{\rm D}(T)} = \cos\left(\lambda\right)\ket{e,\tilde{\al}_{\rm e}(T)} \\ \nonumber&- e^{i(\ww_{\rm q}+ \chi)T}e^{i2F(T)}\sin\left(\lambda\right)e^{-i\left(\ww_{\rm c}-\chi\hat{\sigma}_z\right)\hat{a}^{\dagger}\hat{a}T}\hat{D}(\al_{\rm g}(T))\ket{g,1} \\
&= \cos\left(\lambda\right)\ket{e,\tilde{\al}_{\rm e}(T)} - e^{iG(T)}\sin\left(\lambda\right)\ket{g}\ket{\xi(T)},
\end{align}
where $\tilde{\al}_{\rm e}(T) = \al_{\rm e}(T)e^{-i\left(\ww_{\rm c}+\chi\right)T}$, $G(T) =(\ww_{\rm q}+ \chi)T + 2F(T)$, and $\ket{\xi(T)} = e^{-i\left(\ww_{\rm c}\hat{a}^{\dagger}\hat{a} -\chi\hat{\sigma}_z \hat{a}^{\dagger}\hat{a}\right)T}\hat{D}(\al_{\rm g}(T))\ket{1}$ is the displaced single photon Fock state. Finally, transforming back into the lab frame gives
\begin{flalign}
\nonumber &\ket{\Psi (T)} = \hat{U}^{\dagger}_{\rm D}\left(\cos\left(\lambda\right)\ket{e,\tilde{\al}_{\rm e}(T)} - e^{iG(T)}\sin\left(\lambda\right)\ket{g}\ket{\xi(t)}\right) \\&= \cos\left(\lambda\right)\overline{\ket{e,\tilde{\al}_{\rm e}(T)}} - e^{iG(T)}\sin\left(\lambda\right)\hat{U}^{\dagger}_{\rm D}\ket{g}\ket{\xi(t)}, \label{eqn:eDress}
\end{flalign}
where as before we have made the identification that $\hat{U}^{\dagger}_{\rm D}\ket{e,\tilde{\al}_{\rm e}(T)} = \overline{\ket{e,\tilde{\al}_{\rm e}(T)}}$ to first order in $\lambda$. If $\lambda \ll 1$ such that $\sin(\lambda) \approx 0$, then as before $\ket{\Psi(T)}$ is a dressed coherent state. Additionally, if as an initial state we use $\ket{\Psi(0)} = \overline{\ket{e,0}}$ instead of the bare excited state $\ket{e,0}$, then as $\hat{U}_{\rm D}\overline{\ket{e,0}} = \ket{e,0}$ the term proportional to $\sin(\lambda)$ in equation (\ref{eqn:eDress}) disappears, and $\ket{\Psi (T)}$ contains only the dressed coherent state $\overline{\ket{e,\tilde{\al}_{\rm e}(T)}}$. The dressed initial state is closer to experimental reality as it is an eigenstate of the Hamiltonian, however, the calculation for the bare initial state $\ket{\Psi (0)} = \ket{e,0}$ was shown for completeness, as some experimental protocols can prepare this state, and as from the solution for the bare initial state the solution for the dressed initial state is trivial to obtain.
\newpage
\section{Corrections to the Phase of the Dressed Coherent State Due to Nonlinear Terms}
\label{app:PhaseMatch}

In the expected parameter regime, FIG.~\ref{fig:Overlaps} demonstrates excellent agreement between the state created by a numerical simulation of the Jaynes-Cummings Hamiltonian and the dressed coherent state. However, to obtain this high fidelity, it was necessary to include effects beyond the dispersive Hamiltonian to correctly match the phase of the dressed coherent state amplitude $\tilde{\al}_{\rm g/e}(T)$ with the numerical state. In particular, following Ref.~\cite{Boissonneault:2009kq}, we include the nonlinear term proportional to $\Delta\lambda^4$ in the classical equations of motion for the amplitudes $\tilde{\al}_{\rm g/e}(T)$ and obtain the approximate solutions
\begin{flalign}
&\tilde{\al}_{\rm g}(T) = \al_{\rm g}(T)\exp{\left(-i\left(\ww_{\rm c}-\chi + \zeta\frac{\left<\hat{n}(T)\right>}{2}\right)T\right)}, \\
&\tilde{\al}_{\rm e}(T) = \al_{\rm e}(T)\exp{\left(-i\left(\ww_{\rm c}+\chi - \zeta\left(\frac{\left<\hat{n}(T)\right>}{2}+1\right)\right)T\right)},
\end{flalign}
where $\zeta = \Delta\lambda^4$, $\left<\hat{n}(T)\right> = \left<\hat{a}^\dagger\hat{a}\right>(T)$ is the average photon number in the cavity at time $T$, and $\al_{\rm g/e}(T)$ are defined as before in equation (\ref{eqn:Alphas}). Using these modified coherent state amplitudes in equations (\ref{eqn:apDC}) and (\ref{eqn:eDress}) gives the excellent overlap with the numerical states seen in FIG.~\ref{fig:Overlaps}. We emphasize that the nonlinear corrections have only been used to correct the classical solution for $\tilde{\al}_{\rm g/e}(T)$, and that to lowest order the nonlinearity modifies only the phase and not the amplitude of the dressed coherent states. Furthermore, the squeezing Hamiltonian of the nonlinearity is not considered in our analytical calculations (as this would modify the state so that it was no longer a dressed coherent state), and while its effect is small, we suspect it to be the leading cause of the less than unit fidelity seen in FIG.~\ref{fig:Overlaps}. 

\section{Driven Qubit Excited State Probability}
\label{app:DriveQ}
To first order in the Magnus expansion, the qubit-drive evolution operator is
\begin{align}
\nonumber\hat{U}_{\rm Q}(\tau,0) &= \exp\left\{-i\Omega_1(\tau,0)\right\}\\ 
\nonumber&= \cos\left(\abs{\eta b(\hat{n},\tau)}\left(\nu+2\chi\hat{n}\right)^{-1}\right)\mathbb{I} \\ \nonumber&+ \sin\left(\abs{\eta b(\hat{n},\tau)}\left(\nu+2\chi\hat{n}\right)^{-1}\right)\Big((\eta b(\hat{n},\tau))^*\sigma^{-}\\ &-\eta b(\hat{n},\tau)\sigma^{+}\Big)\abs{\eta b(\hat{n},\tau)}^{-1},
\end{align}
where we have defined the operator function $b(\hat{n},\tau) = 1 - e^{i\left(\nu + 2\chi\hat{n}\right)\tau}$. Starting from the initial state $\ket{\psi'_{\rm D}(0)} = \ket{g,\beta}$ in the dispersive-interaction frame, the state after a time $\tau$ of the applied qubit-drive is
\begin{align}
&\nonumber\ket{\psi'_{\rm D}(\tau)} = \hat{U}_{\rm Q}(\tau,0)\ket{\psi'_{\rm D}(0)} \\ \nonumber&= e^{\frac{-\abs{\beta}^2}{2}}\sum_{k}\frac{\beta^k}{\sqrt{k!}}\Bigg[\cos\left(\frac{\abs{\eta b(k,\tau)}}{\nu+2k\chi}\right)\ket{g,k} \\ &- i \sin\left(\frac{\abs{\eta b(k,\tau)}}{\nu+2k\chi}\right)e^{i\varphi}e^{i\left(\nu + 2k\chi\right)\tau/2}\ket{e,k}\Bigg],
\end{align}
where $e^{i\varphi}=\eta/\abs{\eta}$, and we have used the fact that $b(k,\tau)/\abs{b(k,\tau)} = ie^{i\left(\nu + 2k\chi\right)\tau/2}$. Transforming out of the interaction frame, we arrive at the state in the dispersive frame (after factoring out a global phase)
\begin{align}
\nonumber\ket{\psi_{\rm D}(\tau)} &= \hat{U}^{\dagger}_{\rm I}(\tau,0)\ket{\psi'_{\rm D}(\tau)} \\ \nonumber&= e^{\frac{-\abs{\beta}^2}{2}}\sum_{k}\frac{\tilde{\beta}^k}{\sqrt{k!}}\Bigg[\cos\left(\frac{\abs{\eta b(k,\tau)}}{\nu+2k\chi}\right)\ket{g,k} \\
&- i\sin\left(\frac{\abs{\eta b(k,\tau)}}{\nu+2k\chi}\right)e^{i\varphi}e^{-i\Sigma\tau/2}\ket{e,k}\Bigg],
\end{align}
where $\tilde{\beta} = \beta e^{-i\ww_{\rm c}\tau}$ absorbs the rotation of the cavity state in phase space, and $\Sigma = \ww_{\rm q} + \chi + \ww + 2k\chi = \nu + 2k\chi + 2\ww$. Finally, transforming back into the lab frame, we end up with the state
\begin{flalign}
&\nonumber\ket{\psi_{\rm D}(\tau)} = \hat{U}^{\dagger}_{\rm D}(\tau,0)\ket{\psi_{\rm D}(\tau)} = \\ \nonumber& e^{\frac{-\abs{\beta}^2}{2}}\sum_{k}\frac{\tilde{\beta}^k}{\sqrt{k!}}\Bigg[\cos\left(\frac{\abs{\eta b(k,\tau)}}{\nu+2k\chi}\right)\cos\left(\lambda\sqrt{k}\right)\ket{g,k}  \\ 
\nonumber&-\cos\left(\frac{\abs{\eta b(k,\tau)}}{\nu+2k\chi}\right)\sin\left(\lambda\sqrt{k}\right)\ket{e,k-1} \\ 
\nonumber&- i\sin\left(\frac{\abs{\eta b(k,\tau)}}{\nu+2k\chi}\right)e^{i(\varphi-\Sigma\tau/2)}\cos\left(\lambda\sqrt{k+1}\right)\ket{e,k} \\ 
&- i\sin\left(\frac{\abs{\eta b(k,\tau)}}{\nu+2k\chi}\right)e^{i(\varphi-\Sigma\tau/2)} \sin\left(\lambda\sqrt{k+1}\right)\ket{g,k+1}\Bigg].
\label{eqn:qdriveLab}
\end{flalign}
Now we can calculate the probability at a given time $\tau$ that the qubit is in the excited state, given by ${\rm P}_{e}(\tau) = \bra{\psi_{\rm D}(\tau)}\ket{e}\bra{e}\otimes\mathbb{I}\ket{\psi_{\rm D}(\tau)}$. Using equation (\ref{eqn:qdriveLab}) we calculate this to be
\begin{widetext}
\begin{align}
\nonumber&{\rm P}_{e}(\tau) = e^{-\abs{\beta}^2}\sum_{k}\frac{\abs{\beta}^{2k}}{k!}\Bigg[\cos^2\left(\frac{\abs{\eta b(k,\tau)}}{\nu+2k\chi}\right)\sin^2\left(\lambda\sqrt{k}\right) +\sin^2\left(\frac{\abs{\eta b(k,\tau)}}{\nu+2k\chi}\right)\cos^2\left(\lambda\sqrt{k+1}\right) \Bigg] \\&+2e^{-\abs{\beta}^2}\sum_{k}\frac{\abs{\beta}^{2k}}{k!\sqrt{k+1}}\cos\left(\frac{\abs{\eta b(k+1,\tau)}}{\nu+2(k+1)\chi}\right) \sin\left(\lambda\sqrt{k+1}\right)\sin\left(\frac{\abs{\eta b(k,\tau)}}{\nu+2k\chi}\right)
\cos\left(\lambda\sqrt{k+1}\right)\rm{Im}\left[\tilde{\beta}e^{-i\varphi}e^{i\Sigma\tau/2}\right].
\label{eqn:Eprob} 
\end{align}
\end{widetext}
Equation (\ref{eqn:Eprob}) is quite cumbersome, and to gain some intuitive understanding, we set $\ww = \ww_{\rm q}$ and examine the $\chi,\lambda \rightarrow 0$ limit, which results in equation (\ref{eqn:EprobAp}).

\bibliography{PhaseEffectsL}

\end{document}